\documentclass[10pt, conference]{sig-alternate-10pt}

\usepackage{multirow}
\usepackage{amssymb,amsmath}
\usepackage{graphicx} 
\usepackage{subfig}
\usepackage{algorithm}
\usepackage{algorithmic}
\usepackage{color, soul}
\usepackage{cite}
\usepackage{array}
\usepackage{cases}
\usepackage{url}
\usepackage{colortbl}

\usepackage{multicol}
\usepackage{subfloat}

\definecolor{Gray}{gray}{0.9}

\newcounter{MANumberOfComments}
\stepcounter{MANumberOfComments}

\newcounter{LINumberOfComments}
\stepcounter{LINumberOfComments}

\begin{document}

\title{C3PO: Computation Congestion Control (PrOactive)\\[.3cm]
\Large{- an algorithm for dynamic diffusion of ephemeral in-network services}}

\numberofauthors{4}

\author{
\alignauthor Liang Wang\\
       \affaddr{University of Cambridge, UK}\\
\alignauthor M\'{a}rio Almeida\\
       \affaddr{Telefonica Research, Spain}\\
       \and
\alignauthor Jeremy Blackburn\\
       \affaddr{Telefonica Research, Spain}\\
\alignauthor Jon Crowcroft\\
       \affaddr{University of Cambridge, UK}\\
}

\maketitle

\begin{abstract} 

There is an obvious trend that more and more data and computation are migrating into networks nowadays. Combining mature virtualization technologies with service-centric networking, we are entering into an era where countless services reside in an ISP network to provide low-latency access. Such services are often computation intensive and are dynamically created and destroyed on demand everywhere in the network to perform various tasks. Consequently, these ephemeral in-network services introduce a new type of congestion in the network which we refer to as "computation congestion". The service load need to be effectively distributed on different nodes in order to maintain the functionality and responsiveness of the network, which calls for a new design rather than reusing the centralised scheduler designed for cloud-based services.
In this paper, we study both passive and proactive control strategies, based on the proactive control we further propose a fully distributed solution which is low complexity, adaptive, and responsive to network dynamics.

\end{abstract}

\section{Introduction}
\label{sec:intro}



Looking into the history of computer systems, computation has been migrating between terminal clients and central servers, leading to different designs: "fat client and thin server" or "thin client and fat server". The shifts of this trend are mostly driven by the changes in usage pattern, advances in hardware and software technologies, and even new business models. Nowadays, both clients and servers are rather fat regarding to their processing power and storage capacity, but the fact is that they still fail to keep their pace with the ever-growing demands of end users. Meanwhile, network devices have been quickly evolving and growing their capabilities \cite{telefonica:nfv, telefonica:cartablanco, telefonica:unica}. Quite different from a decade ago, these powerful middle boxes were no longer simple network devices which used to only know how to forward packets. They have complicated structures, highly optimised algorithms, and powerful processing and storage capabilities even comparable to end devices. Since these network devices are mostly underutilised, there is an obvious trend that more and more data and computation are migrating into networks.

Such migration has been accelerated by the following facts in both directions, namely from clouds to networks and from end-user devices to networks. First, many popular Internet services are cloud-based which often rely on a persistent and stable connection to access. However, both connectivity and latency pose significant challenges on quality of services especially in a challenged environment. To improve service availability and reduce latency, big content providers often resort to content-distribution networks (CDN) or deploy their own datacenters co-located with ISP networks. Second, the emergence of User Generated Content (UGC) has further triggered another dramatic shift in usage pattern on the Internet. Huge amount of content is constantly generated and consumed on mobile devices. Processing and storing such overwhelming information, combined with users' increasing on-line activities, give birth to various mobile applications, most of which require a significant amount of computations on users' devices. Given current battery technology, mobile devices are severely energy constrained. Many prior work proposed to offload computation intensive tasks into a network to extend battery life \cite{6195845, Cuervo:2010:MMS:1814433.1814441}.
Third, even for ISP themselves, their own network services started migrating from specialised servers to their networks with the adoption of the NFV (Network function virtualization) paradigm. For example, Telefonica is shifting 30\% of their infrastructure to NFV technologies by the end of 2016\cite{telefonica:nfv, telefonica:cartablanco, telefonica:unica}. Other providers such as AT\&T\cite{att:nfv}, Vodafone\cite{ericsson:nfv}, NTT Docomo\cite{nttdocomo:nfv} and China Mobile\cite{chinamobile:nfv} are following similar strategies.

ISPs' networks, especially those at edges, have transformed into an ideal place for both storing data and performing computation, which collectively provide \textit{services} to its users. Followed by previous information-centric networking (ICN) proposals \cite{jacobson:ccn, Dannewitz:2013:NII:2459510.2459643, 10033435}, service-enabled ICN designs \cite{Nordstrom:2012:SES:2228298.2228308, 6882684, Sathiaseelan:2015:SSC:2753488.2753490, freedman2010service, Braun:2013:SNE:2480362.2480475} clearly start gaining many research interests in the community. Because service execution consumes multiple resources on a router, especially demands CPU cycles for computation intensive tasks, it introduces a new type of "congestion" in a network which we refer to as \textit{"computation congestion"}. Different from conventional traffic congestions which are avoided by the cooperation of both communication ends, in-network services do not necessarily impose a point-to-point paradigm. Also different from classic load balancing problem in cloud which often has a regular structure (i.e., regular network topology, central coordination, homogeneous configurations, uniform demands, and etc.), the situation in an ISP network is more complicated: 1) the underlying topology is not regular; 2) the node configurations can be heterogeneous; 3) demands distribution is highly skewed hence the resources in a neighbourhood needs to be well utilised; 4) central coordination is often expensive and reduces responsiveness of a node.

The emerging ephemeral in-network services call for a thorough investigation on the "computation congestion control" in order to effectively distribute service load within a neighbourhood. In this paper, we study two basic control strategies and propose a fully distributed algorithm called C3PO (Computation Congestion Control PrOactive) built atop of proactive control strategy. Our preliminary evaluations with various realistic settings show that the proposed algorithm is low complexity, able to well exploit neighbourhood resources, and very responsive to dynamic workloads.

\section{Related Work}
\label{sec:related}

ICN is a clean-slate redesign of current Internet to build network infrastructure around content. It abandons the classic point-to-point communication paradigm, and applies two basic design principles in its architecture: 1) accessing content by name and 2) universal caching. Originally, the notion of information in prior ICN proposals\cite{jacobson:ccn, Dannewitz:2013:NII:2459510.2459643, 10033435} only refers to static content. As cloud computing, virtualisation technology become mature enough, more computation are pushed towards edge networks. The definition of information therefore is naturally extended to include both computation and data, which is also referred to as services in most recent work \cite{Nordstrom:2012:SES:2228298.2228308, 6882684, Sathiaseelan:2015:SSC:2753488.2753490, freedman2010service, Braun:2013:SNE:2480362.2480475}. Such service-enabled ICN systems can be considered as an inevitable evolution of ICN paradigm in order to catch up with the growing demands from edge networks and improve quality of service.

Since service execution consumes different resources, both computation and traffic congestions can potentially happen in a network.
Traditional congestion control targets traffic load. The solutions usually either try to reduce the transmission rate or take advantage of multiple paths \cite{Han:2006:MTJ:1217687.1217696, 662909}. In practice all the solutions rely on the cooperation of both ends in a transmission. In ICN context, the congestion needs to be controlled in a hop-by-hop fashion and can be ameliorated by caching to some extent \cite{wangeffects, wong:globecom2012}.

Load balancing, scheduling, and resource management are classic problems in high-performance computing (HPC) cluster. The topic gained lots of attention recently due to the popularity of cloud computing, virtualization, big data framework. Fully centralised control \cite{Hindman:2011:MPF:1972457.1972488, Schwarzkopf:2013:OFS:2465351.2465386} is a popular solution at the moment, and control theory has been shown as an effective tool to dynamically allocate resources \cite{Kalyvianaki:2014:ARP:2642710.2626290}. As mentioned, there are distinctive differences between a cloud environment and an ISP edge network regarding its stability, homogeneous configuration, regular topology, and etc. Most jobs execute for a longer period and often access a lot of data, hence can tolerate long scheduling delay. 

The maturity of virtualisation technologies (e.g., Xen, Linux container, unikernel\cite{Barham:2003:XAV:1165389.945462, Soltesz:2007:COS:1272998.1273025, Madhavapeddy:2013:URV:2557963.2566628}) combined with edge computing will undoubtedly lead us to an era where countless services reside in an ISP network, dynamically created and destroyed on demand to perform various tasks. In such a context, previous highly centralised solution designed for cloud-based services will fail to scale in order to provide a responsive control over such a high volume and asymmetrically distributed demands. Based on our knowledge, very little work has been done to address this challenge. In this paper, we focus on these ephemeral and computation intensive services and research a low complexity, distributed, self-adaptive, and responsive solution.

\section{Proposed Solution}
\label{sec:sol}

We start this section with two fundamental control strategies, followed by a basic workload analysis on a service router, based on which we propose a proactive strategy to avoid computation congestion. Then we present the actual algorithm (C3PO) with implementation details.

\subsection{Two Basic Strategies}
\label{sec:strategy}

Service execution consumes both CPU and memory as well as other resources such as bandwidth. Herein we focus on the first two since they are usually the most dominant resources. The goal of load balancing is achieved by strategically drop or forward the computational tasks to some other nodes to avoid being overloaded. However, instead of distributing load uniformly over all available nodes, a service is preferred to be executed as close to a client as possible to minimise induced latency.

Centralised coordination is not ideal in a practical deployment (especially out of datacenters) due to the obvious reasons: 1) A central solver needs global knowledge of all the nodes in a network; 2) the optimal strategy needs to be calculated periodically given the dynamic nature of a network and traffic; 3) there is a single point of failure; 4) there might be only marginal improvement over a smartly designed heuristic. Therefore, we study two basic strategies in this paper.

\begin{itemize}

\item \textbf{Passive Control}: with this strategy, a node tries to execute as many services as possible before being overloaded. Whenever a service request arrives, it will be executed by default given enough resources. If the node is overloaded, the request will be passed to the next hop along the path to a server, or dropped if current node is already the last hop node in ISP networks.

\item \textbf{Proactive Control}: with this strategy, a node tries to execute services conservatively to avoid being overloaded. To do so, a node estimates request arrival rate with which it can further estimate the potential consumption. If the estimate shows that the node may be overloaded, it only executes some requests and forwards the rest to the next hop neighbour with the lightest load. This strategy requires exchanging state information within a node's one-hop neighbourhood.

\end{itemize}

Because of its simple logic, passive strategy has a very straightforward implementation. Clients can benefit from minimised service latency given no nodes are overloaded, since a service gets executed immediately at an edge router. For proactive strategy, the implementation relies on how estimate is made which we will detail in the following. Despite of being conservative, we still aim to keep the latency low.

\subsection{Workload Analysis}
\label{sec:workload}

A node $n$ receives service requests either from directly connected clients or neighbours. We assume that a node $n$ has CPU capacity $c'$ and memory capacity $m'$.
For a specific service $f_j$, we denote its average CPU and memory consumption as $c_j$ and $m_j$ respectively. In practice, both can be easily measured by tracking a service execution. We also assume the execution time of $f_j$ follows an exponential distribution with its mean value equal to $t_j$. The requests for service $f_j$ can be viewed as a Poisson processes with arrival rate $\lambda_j$. We can easily recognise that the process is a typical \textit{birth-death process}. Because the joint process of multiple Poisson processs is also Poisson, the aggregated requests of all services form another well-defined birth-death process with the birth rate as $\lambda = \sum_{\forall j} \lambda_j$ and death rate as $\mu = \sum_{\forall j} \frac{1}{t_j}$. We herein focus on this aggregate request stream.

In order to calculate average workload, for any given time, we need to estimate the average number of simultaneously running services on node $n$, denoted as $l$. This is equivalent to calculating the average queue length in a simple $M/M/1$ queueing system, where the clients in a queue represents the services running concurrently on a node by applying a multiprogramming model. Herein we consider a stable system where $\lambda < \mu$ to prevent a queue from growing infinitely long to overload a node. We will show later how a proactive strategy is able to keep the system stable.  We have assumed that one CPU is allocated for service execution hence we choose to use $M/M/1$ model in this paper to simplify the discussion. However the analysis can be easily extended to $M/M/C$ model to analyse a multi-core system.

Let $p_j$ denote the normalised popularity of $f_j$ derived from all the requests observed by $n$, then $p_j = \frac{\lambda_j}{\lambda}$ and note that $\sum_{\forall j} p_j = 1$ by definition. The average CPU consumption is $c'' = \sum_{\forall j} p_j \times c''_j$ and average memory consumption is  $m'' = \sum_{\forall j} p_j \times m''_j$. If we let $\rho = \frac{\lambda}{\mu}$ (i.e., utilisation rate), then we have $l = \frac{\rho}{1 - \rho}$ by applying a stationary analysis on $M/M/1$ model. Therefore we can calculate the overall workload induced by executing services in a straightforward way: namely $l \times c''$ for CPU load and $l \times m''$ for memory load.

\subsection{Probabilistic Execution}
\label{sec:prob}

To avoid overloading a node, we need to make sure the workload is less than $n$'s actual capacity. As we have shown, workload is directly controlled by the queue length $l$, which can be further tuned by probabilistically selecting some requests in a stream to execute and forwarding the rest to the next hop. For each service request, if we let node $n$ execute a service with probability $q$, and $q \in [0,1]$ follows a uniform distribution. According to basic queueing theory, the resulting sub-process forms another well-defined birth-death process, with a new birth rate $q \times \lambda$ and the same death rate $\mu$. Therefore the new sub-process has a new utilisation rate equal to $q \times \rho$. To calculate $q$, we can simply perform the following derivations by letting the induced load (e.g., for CPU) $l \times c''$ less than the capacity $c'$.

\begin{align}
l \times c'' < c' & \Longrightarrow \frac{q \times \rho}{1 - q \times \rho} \times c'' < c' \\
& \Longrightarrow \rho \times q < \frac{c'}{c' + c''} \\
& \Longrightarrow  q < \frac{c'}{c' + c''} \times \frac{\mu}{\lambda}
\end{align}

The formula has a very intuitive explanation: if services can be executed faster on average (i.e., higher death rate $\mu$), node $n$ increases $q$ in order to serve more requests by maintaining a longer queue; otherwise $n$ decreases $q$ to reduce the queue length. If requests arrive faster (i.e., higher birth rate $\lambda$), the node also decreases $q$ to keep the number of simultaneously running services low. Similarly, we can perform the same calculations for memory constraint $m'$. Eventually, we set $q$ with the following formula.

\begin{align}
& q = \max\{ \min\{ \frac{c'}{c' + c''}, \frac{m'}{m' + m''} \} \times \frac{\mu}{\lambda}, 1\}
\label{eq:0}
\end{align}

The formula above essentially indicates that the final $q$ is decided by the first bottleneck in a system, either CPU or memory in our case. Also, $q$ is capped by $1$, indicating that an underutilised system will simply accept all the requests. 




\subsection{Proactive Control}
\label{sec:congestion}

We present an implementation of proactive control in Algorithm~\ref{algo:1}, namely \textit{C3PO} -- Computation Congestion Control (PrOactive). The algorithm consists of two major functions: \textbf{on\_arrival($\cdot$)} (line 1--10) is called whenever a service request arrives; and \textbf{on\_complete($\cdot$)} (line 12--21) is called whenever a service execution is completed. The notations used in the algorithm follow the same definition as those in the previous text. By keeping track of CPU usage $c''$, memory usage $m''$, execution rate $\mu$, and request arrival rate $\lambda$, the previous analysis shows how to control the workload by tuning execution probability $q$. However, maintaining a complete history of these statistics can be very expensive. In the actual implementation, we use four circular buffers of size $k$: 1) buf$_\lambda$ for the timestamps of the most recently arrived requests; 2) buf$_\mu$ for the execution time of the most recently finished services; 3) buf$_{c''}$ and 4) buf$_{m''}$ for CPU and memory usage of the most recently finished services.

\begin{algorithm}[!tb]
  \caption{C3PO - A Distributed Proactive Computation Congestion Control for In-Network Services}
  \label{algo:1}
  \begin{algorithmic}[1]
    \STATE{void \textbf{on\_arrival} (request $r$):}

    \STATE{\quad buf$_\lambda$[$i$] $\leftarrow$ timestamp ($r$)}

    \STATE{\quad $\lambda \leftarrow$ mean\_rate (buf$_\lambda$)}
    
    \STATE{\quad $\Delta \lambda \leftarrow$ max$(0, \lambda - \lambda')$}

    \STATE{\quad $\lambda \leftarrow \lambda + \Delta \lambda$}

    \STATE{\quad $q \leftarrow$ \textbf{eq.\ref{eq:0}} ($\lambda, \mu, c', c'', m', m''$)}
    
    \STATE{\quad \textbf{if} draw\_uniform ([0,1]) $ < q$ \textbf{then} execute ($r$)}
    
	\STATE{\quad \textbf{else} forward\_to\_lightest\_load\_node ($r$)}    
    
    \STATE{\quad $i \leftarrow (i+1)$ mod $k$}
    
    \STATE{\quad \textbf{if} $i == 0$ \textbf{then} $\lambda' \leftarrow 0.5 \times $($\lambda' + \lambda - \Delta \lambda$)}

	\STATE
	
	\STATE{void \textbf{on\_complete} (service $s$):}

	\STATE{\quad buf$_\mu$[$i$] $\leftarrow$ execution\_time ($s$)}
	\STATE{\quad buf$_{c''}$[$i$] $\leftarrow$ cpu\_consumption ($s$)}
	\STATE{\quad buf$_{m''}$[$i$] $\leftarrow$ memory\_consumption ($s$)}

	\STATE{\quad $i \leftarrow (i+1)$ mod $k$}
    
    \STATE{\quad \textbf{if} $i == 0$ \textbf{then}}
    \STATE{\quad \quad $\mu \leftarrow 0.5 \times $($\mu +$ mean(buf$_{\mu}$)$^{-1}$)}
    \STATE{\quad \quad $c'' \leftarrow 0.5 \times $($c'' +$ mean (buf$_{c''}$))}
    \STATE{\quad \quad $m'' \leftarrow 0.5 \times $($m'' +$ mean (buf$_{m''}$))}
    
    \STATE{\quad forward\_result (s)}
  \end{algorithmic}
\end{algorithm}

With these four circular buffers, we can calculate the recent values of the parameters in eq.\ref{eq:0}. We decide to use fixed buffer instead of fixed time window to prevent the memory usage of Algorithm \ref{algo:1} from being subject to service arrival/completion rate. Parameter $k$ represents a trade-off between stability and responsiveness. Larger $k$ leads to more stable estimates whereas smaller $k$ indicates higher responsiveness of a strategy to the changes in two metrics (i.e., $\lambda$ and $\mu$). Line 2--6 calculate the execution probability $q$.
The algorithm also maintains a variable $\lambda'$ for the average arrival rate of previous $k$ requests, so that we can calculate the variation in $\lambda$ as $\Delta \lambda = \lambda - \lambda'$. It is definitely worth emphasising line 4 and 5: when $\Delta \lambda > 0$, it indicates an increase in request arrival rate, then C3PO will enter into conservative mode. In conservative mode, C3PO updates $q$ at line 6 by plugging $(\lambda + \Delta \lambda)$ as arrival rate in eq.\ref{eq:0} rather than plugging original $\lambda$. In such a way, C3PO "pessimistically" estimates the arrival rate will increase at the same rate $\Delta \lambda$ in a near future.
If $\Delta \lambda \leq 0$, C3PO operates in normal mode. In some sense, "being proactive" is achieved by "being conservative" when noticing a potential increase in resource consumption.

Although $\lambda$ needs to be calculated at every request arrival (line 3), we can optimise the performance by using another variable $x$ to keep track the sum of arrival intervals.
If we further let $y \leftarrow  \text{buf}_\lambda[(i+1) \text{ mod } k] - \text{buf}_\lambda[i]$ and $z \leftarrow \text{timestamp}(r) - \text{buf}_\lambda[(i-1) \text{ mod } k]$ before performing line 2, then mean rate can be calculated by $\lambda \leftarrow (x - y + z]) / (k - 1) $. Because all $x,y,z$ can be updated with $\mathcal{O}(1)$, this reduces the complexity of "mean\_rate" function from $\mathcal{O}(k)$ to $\mathcal{O}(1)$ by avoiding traversing through all the timestamps in buf$_\lambda$.
Other parameters except $\lambda$ are updated only periodically in both functions (line 10, 18-20). We apply an ARMA (AutoRegressive Moving Average) model with exponential mean when updating these parameters. Both history and recent measure are given the equal weight 0.5. 
To keep the code short and easy to understand, we did not perform further optimisations in Algorithm \ref{algo:1}.

\section{Preliminary Evaluation}
\label{sec:eval}

In our evaluations, we study how different strategy impacts load distribution as well as latency, drop rate, responsiveness to jitters. We test three strategies (None, Passive, and Proactive) on both synthetic and realistic topologies using Icarus simulator \cite{icarus-simutools14}. In most simulations, we use a Poisson request stream with $\lambda = 1000$ as arrival rate, increasing request rate means introducing more load into a network. All simulations are performed at least 50 times to guarantee the reported results are representative.
To simplify the presentation, we assume CPU is the first bottleneck in the system for computation intensive services, and only present the results of using Exodus network \cite{SpringN:Rocketfuel} in the following.

\subsection{Exploiting Neighbourhood}
\label{sec:neighbour}

\begin{figure} [!htp]
\includegraphics[width=8.5cm]{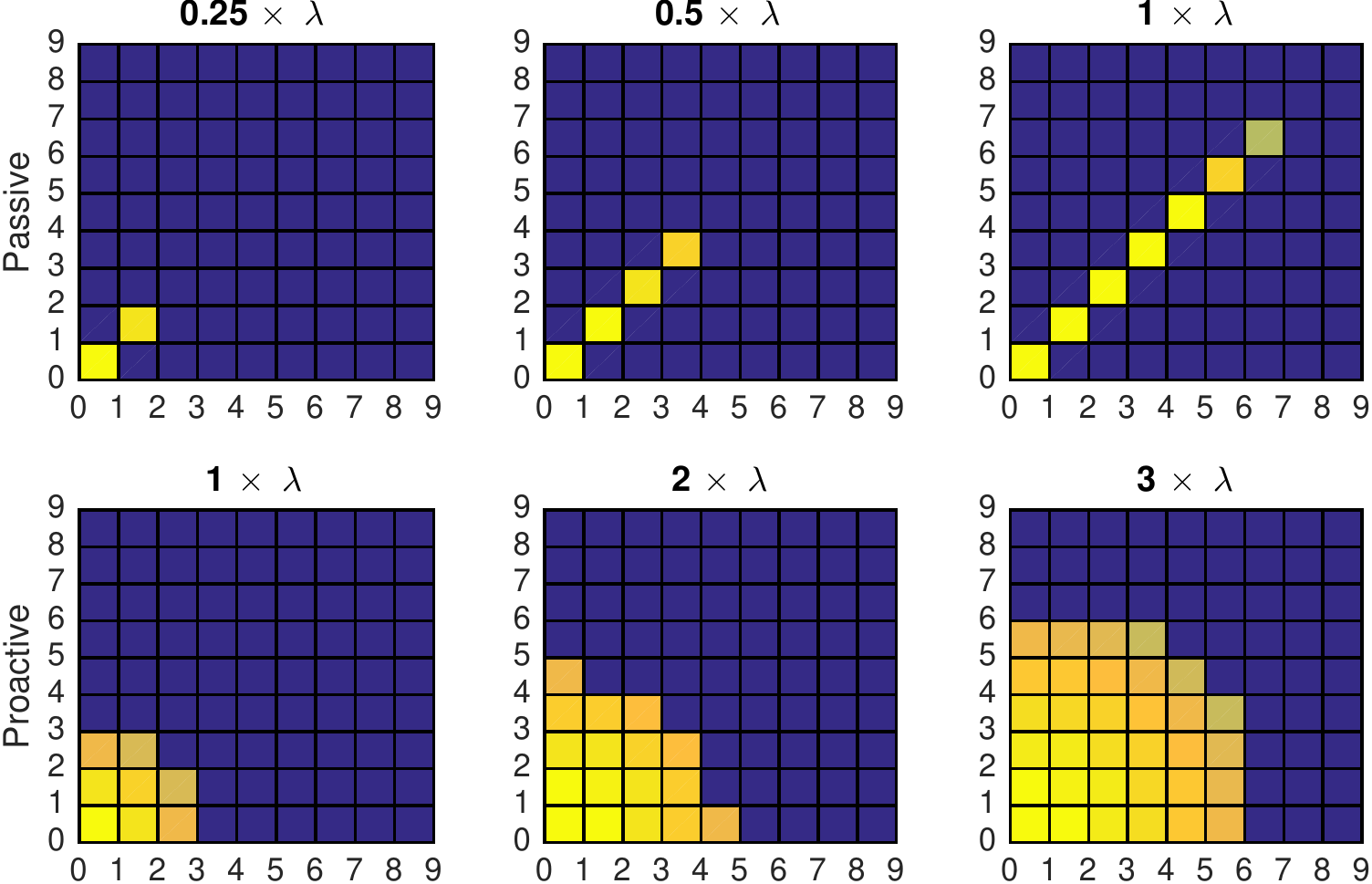}
\caption{An illustration of different behaviours of Passive and Proactive control on grid topology. A client connects to the router at $(0,0)$ while a server connects to the router at $(9,9)$. Proactive is more capable of utilising the nearby resources within its neighbourhood, leading to better load balancing and smaller latency. (Yellow indicates high load.)}
\label{fig:1}
\end{figure}

Before evaluating on a realistic topology, 
figure \ref{fig:1} provides a basic example to illustrate the fundamental differences between passive and proactive strategy. The understanding of these differences will help us in analysing the following results. The experiment is performed on a $10 \times 10$ grid and a router connects to all its adjacent neighbours. For passive control in the first row, since the server is deployed at top right corner, the load is distributed along the path towards the server as we increase the request rate from $0.25 \lambda$ to $\lambda$. Whereas for proactive control, the load is distributed in a quite different way, the services are given high priority to be executed in nearby neighbours. This introduces two immediate benefits: first, a network with proactive control is able to absorb more load. In comparison, with a workload of $3 \lambda$, a large amount of requests will be dropped by the router at (9,9) if passive control is used. Second, because services are likely to be executed on nearby routers, the induced latency tend to be shorter with proactive control. Especially when edge routers are overloaded, the distance between execution point and client grows much slower with proactive control than with passive control as figure shows. Moreover, being able to effectively exploit neighbourhood resources can significantly benefit QoS due to the strong temporal and spatial locality in usage pattern \cite{Wang:2015:PUS:2810156.2810162}.

\subsection{Scalability to Workload}
\label{sec:workload}

\begin{figure} [!htp]
\includegraphics[width=8.5cm]{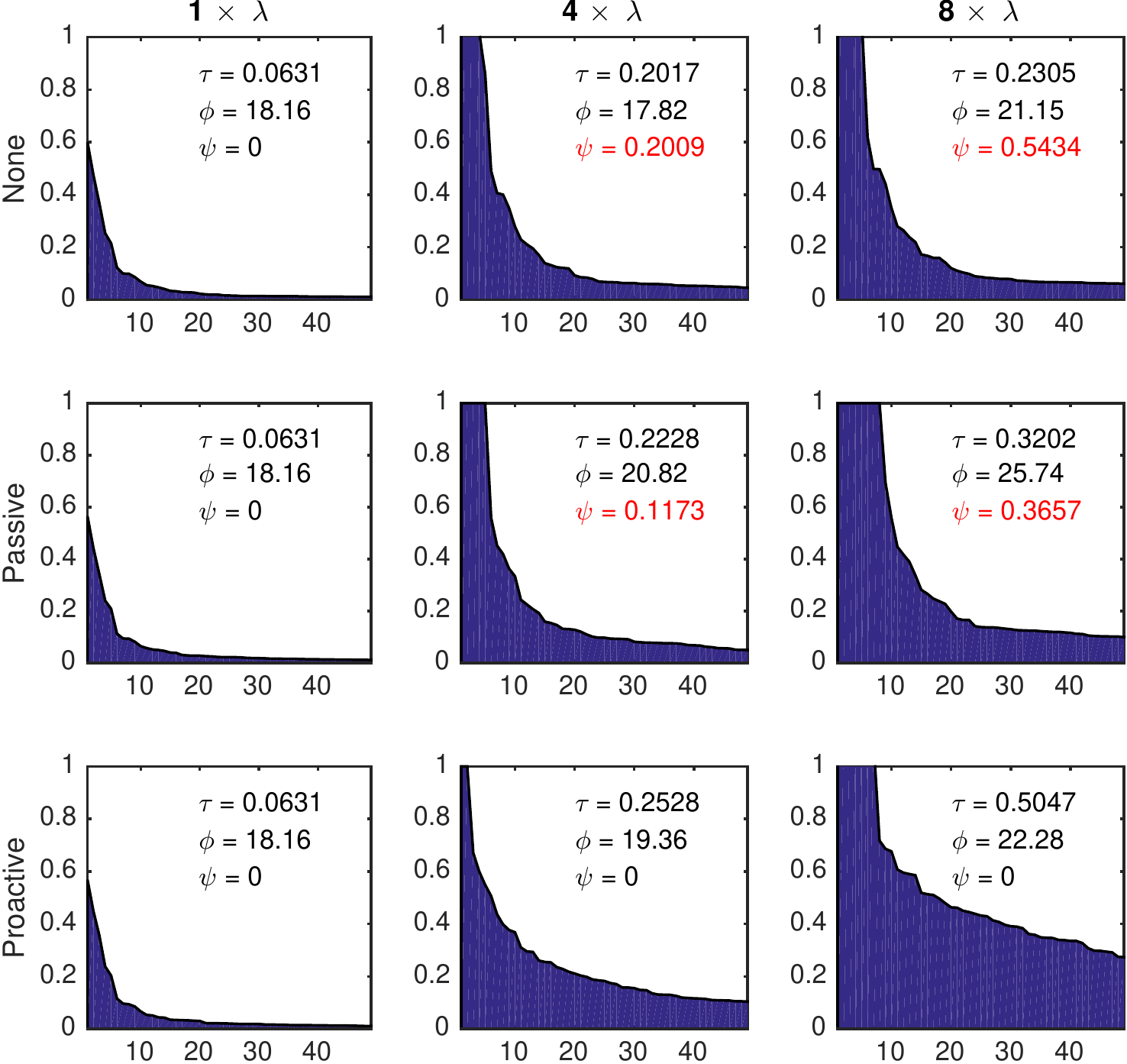}
\caption{Comparison of three control strategies (in each row) on Exodus ISP network, the load is increased step by step in each column. $x$-axis is node index and $y$-axis is load. Top $50$ nodes of the heaviest load are sorted in decreasing order and presented. Notations in the figure: $\tau$: average load; $\phi$: average latency (in $ms$); $\psi$: ratio of dropped requests.}
\label{fig:2}
\end{figure}

Figure \ref{fig:2} shows the results of using three strategies (one for each row) with three workloads (one for each column) on Exodus network. The average load of each node is normalised by its CPU capacity and only top 50 of the heaviest load are presented in a decreasing order in the figure. 

By examining the first column, we can see all three strategies have identical behaviours when the network is underutilised with a workload of $\lambda$. The heaviest loaded node only uses about 60\% of its total capacity. However, as we increase the load to $4 \lambda$ and $8 \lambda$, three strategies exhibit quite different behaviours. For none control at the first row, the figures remain the similar shape. Since no load is distributed and a node simply drops all requests when being overloaded, none control leads to over 54\% drop rate with load of $8 \lambda$. 

For passive control at the second row, we can see both head and tail parts are fatter than none control, indicating more load are absorbed by the network and are distributed on different routers. This can also be verified by checking the average load in the figure: given load $8 \lambda$, passive control increases the average load of the network from $0.2305$ to $0.3202$ comparing to using none control. However, there is still over $36\%$ requests are dropped at the last hop router. This can be explained by the well-known small-world effect which makes the network diameter short in general, so there are only limited resources along a random path.

Among all the experiments, a network with proactive control always absorbs all the load, leading to the highest average load in the network which further indicates the highest utilisation rate. As the workload increases from $\lambda$ to $8 \lambda$, average load also increases accordingly with the same factor. One very distinct characteristic that can be easily noticed in the last two figures on the third row is that the load distribution has a very heavy tail. This is attributed to proactive strategy's capability of offloading services to its neighbours. It is also worth pointing out that we only measured the latency of those successfully executed services, which further explains why none control has the smallest latency, since a service gets executed immediately at an edge router connected to a client, but more than half of the requests are simply dropped and not counted at all. Comparing to passive strategy, proactive strategy achieves shorter latency. Further investigation on other ISP topologies show that such improvement on latency will even increase on larger networks.


\subsection{Responsiveness to Jitters}
\label{sec:jitter}

\begin{figure} [!htp]
\includegraphics[width=9cm]{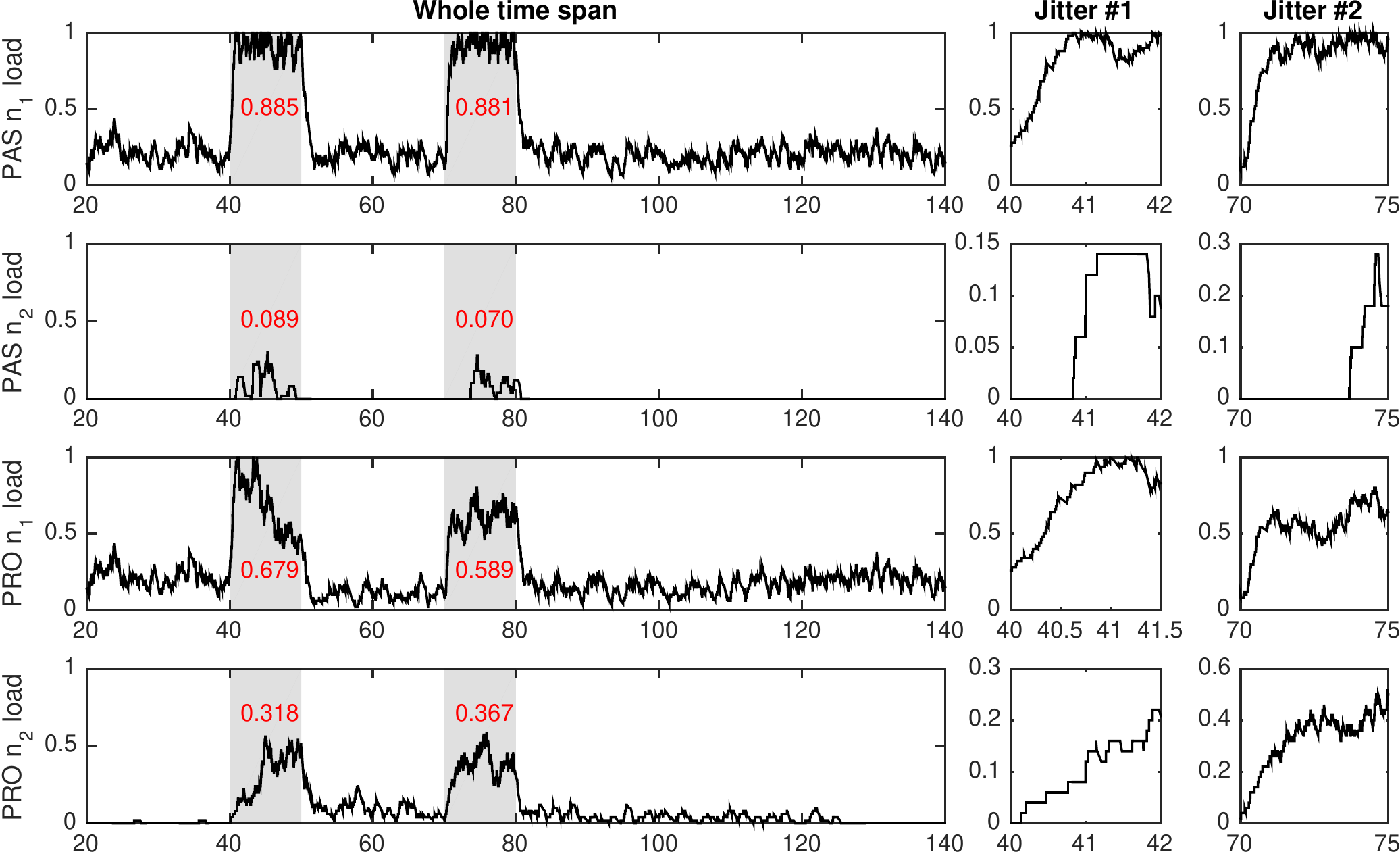}
\caption{Comparison of two control strategies using a simple line topology: client $\rightarrow$ router $n_1$ $\rightarrow$ router $n_2$ $\rightarrow$ server. Two jitters are injected at time 40 ms and 70 ms. $x$-axis is time (ms) and $y$-axis is normalised load. Red numbers represent the average load during a jitter period.}
\label{fig:3}
\end{figure}

To study how a control strategy responds to a sudden increase in workload (a.k.a. jitters), we perform another experiment where we use a simple line topology: client $\rightarrow$ router $n_1$ $\rightarrow$ router $n_2$ $\rightarrow$ server. The client maintains a stable flow of the request rate $\lambda$ and injects two 10-millisecond jitters (of rate $6\lambda$) at time 40 millisecond and 70 millisecond respectively. The first two rows in figure \ref{fig:3} show the time series of the workload on two routers using passive strategy, namely PAS $n_1$ and PAS $n_2$. Similarly, the last two rows are for the two routers using proactive control, namely PRO $n_1$ and PRO $n_2$. The two right columns zoom in at two moments when the jitter just happens (at 40 and 70 ms respectively).

For passive control, the first router PAS $n_1$ takes most of the load (i.e., 88\%) and exhibits consistent behaviours in both jitters. However, the routers using proactive control show an interesting variation when handling two jitters. For the first jitter, although router PRO $n_1$ successfully offloads 31.8\% load to PRO $n_2$, it apparently also experiences high load for a period of 2 ms (i.e., 40 - 42 ms). After the first jitter, PRO $n_1$ enters into a conservative mode, therefore when the second jitter arrives, the load curve on PRO $n_1$ is much flatter and the load peak does not appear at all since it proactively offloads more tasks on PRO $n_2$. As a result, PRO $n_2$ absorbs about 36.7\% load in the second jitter. Even after two jitters, PRO $n_1$ remains in the conservative mode until 130 ms, which explains why there is a small amount of load that has been continuously transferred to PRO $n_2$. After 130 ms, PRO $n_1$ returns to its normal mode. Technically, the mode shift is because all the timestamps of jitters have been purged out from circular buffer buf$_\lambda$ by constant requests.

By checking the second and third columns, we are able to gain an even better understanding on what actually happens when a jitter arrives. For both jitters, proactive control responses faster than the passive one, since the load curve on the second router starts rising earlier and faster. For the second jitter, proactive responses even faster since it is in a conservative mode. Whereas for passive control, PAS $n_2$ only starts taking some load at 74 ms, 4 ms later after the second jitter arrives at PAS $n_1$.

To summarise, our evaluations have clearly showed that proactive control possesses the following attractive properties which make it an ideal solution for balancing computation load in an ISP network: 1) fully distributed with very loose cooperation with one-hop neighbours; 2) good capability of utilising resources in a neighbourhood; 3) high responsiveness to workload jitters.

\section{Conclusion}
\label{sec:conclusion}

We studied and evaluated two control strategies in this paper. Based on the proactive control, we designed a fully distributed, low complexity, and responsive load controller to avoid potential computation congestions when executing in-network services. Our preliminary results showed that the proposed solution C3PO can effectively take advantage of available resources in a neighbourhood to balance the service load and further reduce service latency and request drop rate.

As our future research, we plan to extend the current NDN platform\cite{jacobson:ccn} to implement C3PO. We will perform a more thorough evaluation after a realistic deployment in a production network. Besides, we assumed that the network had enough storage to host all the services. Although in practice, a simple LRU algorithm can be used given a cache is full, a more careful investigation on how caching strategies impact the performance of service execution is definitely needed.

\bibliographystyle{IEEEtran}
\bibliography{references}

\begin{thebibliography}{10}
\providecommand{\url}[1]{#1}
\csname url@samestyle\endcsname
\providecommand{\newblock}{\relax}
\providecommand{\bibinfo}[2]{#2}
\providecommand{\BIBentrySTDinterwordspacing}{\spaceskip=0pt\relax}
\providecommand{\BIBentryALTinterwordstretchfactor}{4}
\providecommand{\BIBentryALTinterwordspacing}{\spaceskip=\fontdimen2\font plus
\BIBentryALTinterwordstretchfactor\fontdimen3\font minus
  \fontdimen4\font\relax}
\providecommand{\BIBforeignlanguage}[2]{{%
\expandafter\ifx\csname l@#1\endcsname\relax
\typeout{** WARNING: IEEEtran.bst: No hyphenation pattern has been}%
\typeout{** loaded for the language `#1'. Using the pattern for}%
\typeout{** the default language instead.}%
\else
\language=\csname l@#1\endcsname
\fi
#2}}
\providecommand{\BIBdecl}{\relax}
\BIBdecl

\bibitem{telefonica:nfv}
{Telefonica\'s view on virtualized mobile networks}. (2015)
  http://www.ict-ijoin.eu/wp-content/uploads/2015/03/6b\_berberana\_telefonica.pdf.

\bibitem{telefonica:cartablanco}
{Carta Blanco: NFV at Telefonica}. (2014)
  http://telecoms.com/interview/carta-blanco-nfv-in-telefonica/.

\bibitem{telefonica:unica}
{Telefonica selects Ericsson for global UNICA program}. (2016)
  http://www.ericsson.com/news/1988285.

\bibitem{6195845}
S.~Kosta, A.~Aucinas, P.~Hui, R.~Mortier, and X.~Zhang, ``Thinkair: Dynamic
  resource allocation and parallel execution in the cloud for mobile code
  offloading,'' in \emph{INFOCOM, 2012 Proceedings IEEE}, March 2012, pp.
  945--953.

\bibitem{Cuervo:2010:MMS:1814433.1814441}
E.~Cuervo, A.~Balasubramanian, D.-k. Cho, A.~Wolman, S.~Saroiu, R.~Chandra, and
  P.~Bahl, ``Maui: Making smartphones last longer with code offload,'' in
  \emph{Proceedings of the 8th International Conference on Mobile Systems,
  Applications, and Services}, ser. MobiSys '10.\hskip 1em plus 0.5em minus
  0.4em\relax New York, NY, USA: ACM, 2010, pp. 49--62.

\bibitem{att:nfv}
{AT\&T Domain 2.0 Vision White Paper}. (2013)
  https://www.att.com/common/about\_us/pdf/at\&t

\bibitem{ericsson:nfv}
{Ericsson and Vodafone deploy first cloud-based VoLTE}. (2014)
  http://www.ericsson.com/news/1993653.

\bibitem{nttdocomo:nfv}
{DOCOMO Partners with Ericsson, Fujitsu and NEC for NFV Deployment}. (2015)
  https://www.nttdocomo.co.jp/english/info/\\media\_center/pr/2015/0302\_00.html.

\bibitem{chinamobile:nfv}
{Alcatel-Lucent and China Mobile conduct industry-first live field trial of a
  virtualized radio access network}. (2015)
  https://www.alcatel-lucent.com/press/2015/alcatel-lucent-and-china-mobile-conduct-industry-first-live-field-trial-virtualized-radio-access.

\bibitem{jacobson:ccn}
V.~Jacobson, D.~K. Smetters, J.~D. Thornton, M.~F. Plass, N.~H. Briggs, and
  R.~L. Braynard, ``Networking named content,'' in \emph{Proceedings of the 5th
  ACM Conext}.\hskip 1em plus 0.5em minus 0.4em\relax New York, NY, USA: ACM,
  2009, pp. 1--12.

\bibitem{Dannewitz:2013:NII:2459510.2459643}
\BIBentryALTinterwordspacing
C.~Dannewitz, D.~Kutscher, B.~Ohlman, S.~Farrell, B.~Ahlgren, and H.~Karl,
  ``Network of information (netinf) - an information-centric networking
  architecture,'' \emph{Comput. Commun.}, vol.~36, no.~7, pp. 721--735, Apr.
  2013. [Online]. Available:
  \url{http://dx.doi.org/10.1016/j.comcom.2013.01.009}
\BIBentrySTDinterwordspacing

\bibitem{10033435}
D.~Trossen, G.~Parisis, K.~Visala, B.~Gajic, J.~Riihijarvi, P.~Flegkas,
  P.~Sarolahti, P.~Jokela, X.~Vasilakos, C.~Tsilopoulos, and S.~Arianfar,
  ``{PURSUIT Conceptual Architecture: Principles, Patterns and sub-Components
  Descriptions},'' PURSUIT, Tech. Rep., May 2011.

\bibitem{Nordstrom:2012:SES:2228298.2228308}
E.~Nordstr\"{o}m, D.~Shue, P.~Gopalan, R.~Kiefer, M.~Arye, S.~Y. Ko,
  J.~Rexford, and M.~J. Freedman, ``Serval: An end-host stack for
  service-centric networking,'' in \emph{Proceedings of the 9th USENIX
  Conference on Networked Systems Design and Implementation}, ser.
  NSDI'12.\hskip 1em plus 0.5em minus 0.4em\relax Berkeley, CA, USA: USENIX
  Association, 2012, pp. 7--7.

\bibitem{6882684}
D.~Griffin, M.~Rio, P.~Simoens, P.~Smet, F.~Vandeputte, L.~Vermoesen,
  D.~Bursztynowski, and F.~Schamel, ``Service oriented networking,'' in
  \emph{Networks and Communications (EuCNC), 2014 European Conference on}, June
  2014, pp. 1--5.

\bibitem{Sathiaseelan:2015:SSC:2753488.2753490}
A.~Sathiaseelan, L.~Wang, A.~Aucinas, G.~Tyson, and J.~Crowcroft, ``Scandex:
  Service centric networking for challenged decentralised networks,'' in
  \emph{Proceedings of the 2015 Workshop on Do-it-yourself Networking: An
  Interdisciplinary Approach}, ser. DIYNetworking '15.\hskip 1em plus 0.5em
  minus 0.4em\relax New York, NY, USA: ACM, 2015, pp. 15--20.

\bibitem{freedman2010service}
M.~J. Freedman, M.~Arye, P.~Gopalan, S.~Y. Ko, E.~Nordstrom, J.~Rexford, and
  D.~Shue, ``Service-centric networking with scaffold,'' DTIC Document, Tech.
  Rep., 2010.

\bibitem{Braun:2013:SNE:2480362.2480475}
T.~Braun, A.~Mauthe, and V.~Siris, ``Service-centric networking extensions,''
  in \emph{Proceedings of the 28th Annual ACM Symposium on Applied Computing},
  ser. SAC '13.\hskip 1em plus 0.5em minus 0.4em\relax New York, NY, USA: ACM,
  2013, pp. 583--590.

\bibitem{Han:2006:MTJ:1217687.1217696}
\BIBentryALTinterwordspacing
H.~Han, S.~Shakkottai, C.~V. Hollot, R.~Srikant, and D.~Towsley, ``Multi-path
  tcp: A joint congestion control and routing scheme to exploit path diversity
  in the internet,'' \emph{IEEE/ACM Trans. Netw.}, vol.~14, no.~6, pp.
  1260--1271, Dec. 2006. [Online]. Available:
  \url{http://dx.doi.org/10.1109/TNET.2006.886738}
\BIBentrySTDinterwordspacing

\bibitem{662909}
L.~Vicisano, J.~Crowcroft, and L.~Rizzo, ``Tcp-like congestion control for
  layered multicast data transfer,'' in \emph{INFOCOM '98. Seventeenth Annual
  Joint Conference of the IEEE Computer and Communications Societies.
  Proceedings. IEEE}, vol.~3, Mar 1998, pp. 996--1003 vol.3.

\bibitem{wangeffects}
L.~Wang, S.~Bayhan, and J.~Kangasharju, ``Effects of cooperation policy and
  network topology on performance of in-network caching,'' \emph{IEEE
  Communication Letters}, 2014.

\bibitem{wong:globecom2012}
W.~Wong, L.~Wang, and J.~Kangasharju, ``{Neighborhood Search and Admission
  Control in Cooperative Caching Networks},'' in \emph{Proc. of IEEE GLOBECOM},
  December 3-7 2012.

\bibitem{Hindman:2011:MPF:1972457.1972488}
B.~Hindman, A.~Konwinski, M.~Zaharia, A.~Ghodsi, A.~D. Joseph, R.~Katz,
  S.~Shenker, and I.~Stoica, ``Mesos: A platform for fine-grained resource
  sharing in the data center,'' in \emph{Proceedings of the 8th USENIX
  Conference on Networked Systems Design and Implementation}, ser.
  NSDI'11.\hskip 1em plus 0.5em minus 0.4em\relax Berkeley, CA, USA: USENIX
  Association, 2011, pp. 295--308.

\bibitem{Schwarzkopf:2013:OFS:2465351.2465386}
M.~Schwarzkopf, A.~Konwinski, M.~Abd-El-Malek, and J.~Wilkes, ``Omega:
  Flexible, scalable schedulers for large compute clusters,'' in
  \emph{Proceedings of the 8th ACM European Conference on Computer Systems},
  ser. EuroSys '13.\hskip 1em plus 0.5em minus 0.4em\relax New York, NY, USA:
  ACM, 2013, pp. 351--364.

\bibitem{Kalyvianaki:2014:ARP:2642710.2626290}
E.~Kalyvianaki, T.~Charalambous, and S.~Hand, ``Adaptive resource provisioning
  for virtualized servers using kalman filters,'' \emph{ACM Trans. Auton.
  Adapt. Syst.}, vol.~9, no.~2, pp. 10:1--10:35, Jul. 2014.

\bibitem{Barham:2003:XAV:1165389.945462}
P.~Barham, B.~Dragovic, K.~Fraser, S.~Hand, T.~Harris, A.~Ho, R.~Neugebauer,
  I.~Pratt, and A.~Warfield, ``Xen and the art of virtualization,''
  \emph{SIGOPS Oper. Syst. Rev.}, vol.~37, no.~5, pp. 164--177, Oct. 2003.

\bibitem{Soltesz:2007:COS:1272998.1273025}
S.~Soltesz, H.~P\"{o}tzl, M.~E. Fiuczynski, A.~Bavier, and L.~Peterson,
  ``Container-based operating system virtualization: A scalable,
  high-performance alternative to hypervisors,'' \emph{SIGOPS Oper. Syst.
  Rev.}, vol.~41, no.~3, pp. 275--287, Mar. 2007.

\bibitem{Madhavapeddy:2013:URV:2557963.2566628}
A.~Madhavapeddy and D.~J. Scott, ``Unikernels: Rise of the virtual library
  operating system,'' \emph{Queue}, vol.~11, no.~11, pp. 30:30--30:44, Dec.
  2013.

\bibitem{icarus-simutools14}
L.~Saino, I.~Psaras, and G.~Pavlou, ``Icarus: a caching simulator for
  information centric networking (icn),'' in \emph{Proceedings of the 7th
  International ICST Conference on Simulation Tools and Techniques}, ser.
  SIMUTOOLS '14.\hskip 1em plus 0.5em minus 0.4em\relax ICST, Brussels,
  Belgium, Belgium: ICST, 2014.

\bibitem{SpringN:Rocketfuel}
N.~Spring, R.~Mahajan, and D.~Wetherall, ``Measuring {ISP} topologies with
  rocketfuel,'' in \emph{Proceedings of ACM SIGCOMM}, 2002.

\bibitem{Wang:2015:PUS:2810156.2810162}
L.~Wang, S.~Bayhan, J.~Ott, J.~Kangasharju, A.~Sathiaseelan, and J.~Crowcroft,
  ``Pro-diluvian: Understanding scoped-flooding for content discovery in
  information-centric networking,'' in \emph{Proceedings of the 2Nd
  International Conference on Information-Centric Networking}, ser. ICN
  '15.\hskip 1em plus 0.5em minus 0.4em\relax New York, NY, USA: ACM, 2015, pp.
  9--18.

\end{thebibliography}

\end{document}